\documentclass[manuscript]{aastex} 
\newcommand{\etal}{{\it et al.\/}}

\begin{document}
\pagenumbering{arabic}

\title{THE EVOLUTION OF GALAXY MORPHOLOGY FOR GALAXIES IN THE\\
CANADA-FRANCE REDSHIFT SURVEY}
 
\author{Sidney van den Bergh}
\affil{Dominion Astrophysical Observatory, Herzberg Institute of Astrophysics, 
National Research Council of Canada, 5071 West Saanich Road, Victoria, BC, 
Canada  V9E 2E7}
\email{sidney.vandenbergh@nrc.ca}

\begin{abstract}

The images of 229 galaxies in the Canada France Redshift Survey have been 
classified on the DDO system. These observations were combined with previous 
classifications of galaxies with known redshifts in the Hubble Deep Field. The 
combined sample provides homogeneous morphological classifications for 425 
galaxies of known redshift. The fraction of all galaxies that are of types E + 
S0 + E/S0 appears to remain approximately constant at $\sim 17\%$ over the 
redshift range $0.25 < z < 1.2$. Over the same range the fraction of irregular 
(Ir) galaxies increases from $\sim 5\%$ to $\sim 12\%$. Part of this increase 
may be due to mild luminosity evolution of Ir galaxies. The frequency of mergers 
is found to rise by a factor of two or three over the redshift range covered by 
the present survey. These results are in qualitative agreement with those 
obtained previously by \citet{binchmannetal1998} using a coarser galaxy 
classification system.
\end{abstract}

\section{INTRODUCTION}

The Canada-France Redshift Survey (CFRS) provides a large well-defined sample 
\citep{lilly1995b,crampton1995} of faint galaxies for which the selection 
criteria match as closely as possible those for samples of nearby galaxies. The 
galaxies in the CFRS have a median redshift of $z \sim 0.6$, which corresponds 
to a look-back time of half the present age of the Universe. The sample gives 
information on the evolution of galaxy morphology over the range $0.0 < z < 1.2$ 
\citep{binchmannetal1998}. It therefore supplements similar studies of the 
evolution of galaxy morphology in the Hubble Deep Field (HDF) 
\citep{williamsetal1996} and its Flanking Fields (FF) by \citet{vdb2000} and by 
\citet{vdbcohencrabbe2001}. Such additional information is of value because (1) 
the numbers of galaxies in individual classification bins is small, and (2) the 
distribution of classification types along any particular line of sight may be 
affected by the presence of voids or density enhancements.

\section{GALAXY CLASSIFICATIONS}

A catalog of classifications on the DDO System 
\citep{vdb1960a,vdb1960b,vdb1960c} of 229 galaxies for which redshifts have been 
published by \citet{LeFevreetal1995, lilly1995a, lilly1995b, hammer1995, 
binchmannetal1998} is given in Table 1. Not listed in Table 1 are 17 objects 
with redshift $z=0.0000$ which were mostly classified as E0, E0/Star or Star. 
Exceptions are CFRS 10.0808, which was classified as Peculiar, and CFRS 10.1614 
which is a distant Sab/S0: galaxy, even though its redshift is claimed to be 
0.0000 on the basis of the resemblance of its spectrum to that of an M-type 
star. Infrared F814W images obtained with the Hubble Space Telescope Wide Field 
Planetary Camera 2 were available for all of the classified galaxies. For many 
of these galaxies a red F606W image had also been obtained. Finally a few blue 
F450W images were also available. Since these blue images were generally 
``underexposed'' they were mostly only useful to show the distribution of the 
young population component in program galaxies. Inspection of (and 
classification from) the F814W and F606W images showed that, for galaxies with 
$0.0 < z < 1.1$, the classifications were only weakly dependent on the rest- 
wavelength at which they were being viewed. The classification effects were in 
the sense that objects viewed at a rest-wavelength longer than that of the  B 
band ($\sim4400 \AA$) appeared to have bulges that were too large and spiral 
arms that were too weak. Only for a few galaxies at $z  > 1.0$ are morphological 
classifications expected to be seriously affected by young blue stars. To 
minimize the effects of rest-wavelength on the classifications, all galaxies 
with $z > 0.60$ were classified on the F814W images. Galaxies with $0.10 < z < 
0.60$ where, whenever possible, classified on F606W images. A small number of 
objects with $z < 0.10$ were classified on the F450W images. Galaxies which were 
classified at (or close to) the rest-frame B wavelength are marked by an 
asterisk in the fourth colum of Table 1. Uncertain classifications are followed 
by a colon. The exposure times for the F814W images used in the CFRS survey 
ranged from 5300 to 7800 seconds. The corresponding exposure times of F814W 
images in the Hubble Deep Field \citep{williamsetal1996}  were 124000 seconds. 
As a result of this difference in exposure times the CFRS images are noisier 
than the HDF images. This difference makes the classification of the CFRS images 
somewhat more challenging than that of the galaxies in the HDF. However, it is 
our impression that this difference is not likely to introduce significant 
systematic differences between classifications in the CFRS and in the HDF. By 
introducing artificial noise \citet{vdbcohencrabbe2001} have found that such 
noise does not significantly influence the classifications of the main bodies of 
galaxies. However, the visibility of dim outer features, such as faint outer 
tidal structure, does depend on the noise level in each image. As was the case 
for the previous classifications of objects in the HDF, it was usually not 
possible to distinguish with confidence between elliptical (E) and lenticular 
(S0) galaxies. Furthermore a few distant cD galaxies might have been 
misclassified as objects of types E-Sa-Sab. For distant spirals it is often 
difficult to use spiral arm morphology as a classification criterion. This is so 
because of resolution effects on the tightly coiled arms of early-type galaxies, 
and because the arms of distant late-type galaxies often appear to have a rather 
chaotic structure. Central concentration of light has therefore been used as the 
principal criterion for locating an individual object on Hubble's (1936) Sa - Sb 
- Sc sequence. Those galaxies with the largest nuclear bulges were assigned to 
type Sa, those with somewhat smaller bulges were classified as being of type Sb, 
and spirals exhibiting the smallest bulges were designated Sc. Finally, galaxies 
lacking a nucleus or nuclear bulge were generally classified as irregulars. All 
classifications were made ``blind'', i.e. without knowing what the redshift of 
that galaxy was.

\section{COMPARISON WITH PREVIOUS CLASSIFICATIONS}

\citet{binchmannetal1998}  previously classified many of the galaxies in the 
present program on the sequence: 1 = E + E/S0, 2 = S0/Sa, 3 = Sab, 4 = Sbc, 5 = 
Scd, and 6= Ir. Their classifications were made with respect to nine fundamental 
types that are illustrated in their Figure 4 \footnote{I am indebted to Simon 
Lilly for pointing out that the galaxies reproduced in this figure by Binchmann 
et al. should only be seen as illustrative examples, and must not be regarded as 
exact prototypes for the morphological classes that they represent.} After 
excluding star-like objects, and those classifications marked ``?'', there were 
147 galaxies that that had been classified 1 to 6 by Binchmann et al. and E-
E/Sa-Sa-Sab-Sb-Sbc-Sc-Sc/Ir-Ir in Table 1. After translating our classifications 
into the slightly coarser sequence
of Binchmann et al. \footnote{The adopted transformation was E + E/S0 = 1, S0/a 
= 2, Sa = 2.5, Sab = 3, Sb = 3.5, Sbc = 4, Sc = 4.5, Scd = 5 and Ir = 6.}   it 
is found (see Figure 1) that 79/147 (54\%) of the classifications differ by 0.5 
units or less. Such close agreement is considered satisfactory because it is 
often difficult to fit the morphologies of distant galaxies into the 
classification bins of the Hubble (1936) scheme \citep{vdb1996,vdb2000}. 
Furthermore inspection of Figure 1 shows no evidence for a significant 
systematic difference between the present classifications on the DDO system and 
those by Binchmann et al. It should, however, be noted that quite a few galaxies 
that Binchmann et al. classify as 6 (irregular) were regarded as mergers in the 
present investigation. In most of these cases careful inspection of the images 
shows the presence of two (or more) nuclei. The best example of an interacting 
pair that was classified as an irregular is CFRS 14.1129 (see Figure 2). This 
object seems to consist of a late-type spiral that is being tidally distorded by 
a compact early-type companion.

\section{DISCUSSION}  
      
\subsection {The distance distribution of sample galaxies}

Figure 3 shows a histogram of the distribution in redshift of the galaxies in 
the present sample. The figure shows an increase by about 40\% in the number of 
galaxies per redshift bin between z = 0.0 and z = 1.0, and then a precipitous 
drop, which is no doubt due to incompleteness of our sample, beyond $z \sim 
1.0$. The fact that the present sample was drawn from objects located in four 
different fields reduces the effects of galaxy clustering along the line of 
sight to any individual field. Figure 4 compares the distribution of spiral and 
elliptical galaxies. Inspection of this figure suggests, and a Kolmogorov-
Smirnov test confirms, that these two distributions do not differ at a 
respectable level of statistical significance. Inspection of this figure 
suggests, and a Kolmogorov-Smirnov test confirms, that these two distributions 
do not differ at a respectable level of statistical significance. By the same 
token it is found that the redshift distributions of the 35 E galaxies and of 
19.5 spirals of type Sc in the CFRS sample do not differ at a statistically 
significant level.

\subsection{Merger frequency as a function of redshift}

In a previous paper \citep{vdb2000} it was found that a number of the objects 
that had tentatively been designated as ``mergers'' had significantly different 
radial velocities and were therefore close optical pairs. It was therefore 
decided to tighten the definition of mergers by restricting it to galaxies that 
exhibited clear evidence for tidal distortion, or to close pairs that are 
embedded in a common envelope. In agreement with previous work  (see Carlberg et 
al. 2000 for a recent review) it is found that the fraction of merging galaxies 
is highest at the largest redshifts. In the present sample 9/22 (41\%) of all 
merging galaxies are located at $z\ge 0.80$, whereas only 38/203 (19\%) of non-
merging galaxies have $z \ge 0.80$.

\subsection{The frequency of barred spirals}

Van den Bergh {\it et al.} (1996) found that barred spirals were unexpectedly 
rare in the Hubble Deep Field North. Subsequently \citet{abrahametal1999} noted 
a striking decline in the fraction of barred spirals beyond redshifts of 0.5-0.7 
in the Hubble Deep Field South. Furthermore observations at a fixed rest-frame 
wavelength \citep{vdb2000} appear to show that this deficiency of barred spirals 
at large redshifts can not, as had been suggested by \citet{eskridgeetal2000}, 
be attributed to wavelength dependent visibility of bars. Van den Bergh {\it et 
al.} (2000) found that only 6\% of 49 galaxies with $0.25 < z < 0.60$ had bars 
or possible bars. Furthermore only 1\% of 134 galaxies with redshifts in the 
range $0.6 < z < 1.2$ appeared to exhibit bar-like structure. [The single barred 
spiral at z = 0.96 might actually be a tidally deformed normal spiral.] For 
comparison it is noted that 22\% of the nearby galaxies in the Shapley Ames 
Catalog \citet{sandagetammann1981} are listed as being barred objects of types 
SB or S(B). It is presently being investigated (Abraham \& van den Bergh in 
preparation) to which extent the fraction of galaxies that is recognized as 
being barred spirals might depend on redshift via resolution and noise-dependent 
effects. In the present study special attention was paid to the presence (or 
absence) of bars, by examining each galaxy image over the widest possible 
dynamical range. It is therefore surprising that only 14 out of 225 (6\%) 
galaxies in the CFRS sample are seen to have bars. Unfortunately the present 
sample is too small to establish if the fraction of barred galaxies decreases 
still more towards increasing redshift.

\subsection{Frequency of grand-design spirals}

Galaxies of DDO luminosity classes I and II exhibit beautifull long arms and are 
commonly denoted as ``grand design'' spirals. Inspection of SRC Schmidt plates 
\citep{visvanathanvdb1992} shows that such spiral arms can still be recognized 
on the tiny Schmidt images of galaxies with redshifts of 10 000 $< v <$ 15 000 
km s$^{-1}$. Such arms are also clearly seen in the reproduction of galaxy CFRS 
10.0826 at z= 0.64, which Binchmann et al. use as their prototype for galaxies 
of type Scd. The present sample contains only six grand-design spirals, three of 
wich have redshifts $z < 0.3$. Our sample is therefore too small to study the 
redshift distribution on grand-design spirals.

\subsection{Evolutionary changes in galaxy morphology}

The relative frequencies of differing morphological galaxy types in different 
redshift ranges are listed in Table 2. The interpretation of these data is 
rendered particularly uncertain by the fact that 185 galaxies are sorted into 
nine morphological bins in three redshift ranges. On average each bin in the 
table therefore contains only $\sim 7$ (185/27) galaxies! The only relatively 
secure conclusion to be drawn from the data in Table 2 is that the total number 
of objects classified as Peculiar,``?'', Protogalaxy and Merger appears to 
increase with redshift. Also given in Table 2 is a comparison between the 
relative frequencies of various galaxy classification types in the present CFRS 
sample and in the Hubble Deep Field plus Flanking Fields studied by 
\citet{vdb2000}. No gross differences appear to show up between the frequency 
distributions of morphological types in these two samples.

\section{COMBINED CFRS AND HDF SAMPLES}

Since the present data on CFRS galaxies have been classified on the same system 
as those in the Hubble Deep Field \citep{vdb2000} the CFRS and the HDF+FF 
samples may be combined. This has two advantages. In the first place the size of 
the sample is approximately doubled. Secondly combining samples observed along 
different lines of sight reduces the bias that could be introduced by the fact 
that a particular line of sight might pass through an overdensity, in which 
early-type galaxies are expected to be over-represented \citep{dressler1980}. 
Data on the resulting frequency distribution of morphological types of 425 
galaxies with $0.25 < z < 1.20$ is collected in Table 3. Subject to the 
inevitable limitations of small number statistics it appears that the following 
tentative conclusions may be drawn from the data in this table:

\begin{itemize}
\item The fraction of E + S0 + E/S0 galaxies remains approximately constant at 
$\sim17\%$ over the range z = 0.25 to z = 1.20.

\item The fraction of irregular galaxies appears to increase slightly from $\sim 
5\%$ at small redshifts to $\sim10\%$ at z $\sim$1. This is the opposite of what 
might have been expected from observational selection. This is so because Ir 
galaxies are, on average, significantly fainter than those of Hubble types E-Sa-
Sb-Sc (see Fig.1 of van den Bergh 1998). This suggests that either (1) irregular 
galaxies were intrinsically more frequent in the distant past than they are now, 
or (perhaps more plausibly) (2) that Ir galaxies experienced more luminosity 
evolution than did objects of types E-Sa-Sb-Sc. However, some constraints on 
such luminosity evolution are set by the observation \citep{carollolilly2000} 
that that star forming galaxies at $0.5 < z < 1.0$ do not appear to have the 
extremely low metallicities that are diagnostic of dwarf galaxies.
 
\item The frequency of galaxy mergers appears to increase by a factor of two or 
three over the range $0.4 < z < 1.0$. A significant fraction of the distant 
objects, that had previously been classified as irregulars, turn out to be 
mergers.

\end{itemize}
 
The present investigation suggests that larger galaxy samples, that are drawn 
from separate fields in the sky, will be needed to derive statistically 
significant conclusions on the redshift dependence of (1) the merger rate, (2) 
the fraction of grand-design spirals and (3) the fraction of barred spirals. It 
is a pleasure to thank Simon Lilly for kindly providing the HST images of CFRS 
galaxies and to Peter Stetson for his help with help with the reduction of these 
images. I also thank Roberto Abraham and Simon Lilly for critical reading of an 
early draft of this paper.

\clearpage

\begin{deluxetable}{lllll}
\tablewidth{470pt}
\tablecaption{\sc{Morphological classifications for galaxies that have been 
observed spectroscopically.} }
\tablehead{\colhead{$ z $} & \colhead{CFRSNo.} & \colhead{Classification} & 
\colhead{ *?} & \colhead{Remarks}}
\startdata

$0.0000$ & $10.1502$ & Sab/S0:	& 	& galaxy morphology shows  redshift is 
in error\\
$0.0067$ & $10.1650$ & Ir V:	& 	& nearby dwarf irregular\\
$0.0000$ & $10.1502$ & Sab/S0:	& 	& galaxy morphology shows that 
redshift is in error\\
$0.0067$ & $10.1650$ & Ir V:	& 	&nearby dwarf irregular\\
$0.0330$ & $03.0949$ & Sc	&*	&edge-on\\
$0.0640$ & $14.0528$ & E2/Merger& 	&has binary nucleus\\
$0.0684$ & $14.0435$ & Pec	& 	&nuclear regions contains dust and star 
formation,\\
         &           &          &       & which is embedded in an E4-like 
envelope\\
$0.0730$ & $10.1612$ & Amorphous? & 	&\\	
$0.0740$ & $10.0767$ & Ba?	  & 	&damaged image or contains dust\\
$0.0760$ & $10.1613$ & Pec        & 	&\\		
$0.0767$ & $10.1664$ & S pec	  & 	&\\
$0.0781$ & $14.1193$ & Tadpole    & 	&\\
$0.0793$ & $14.1039$ & dIr	  & 	&edge-on\\
$0.0810$ & $14.0685$ & E3	  &	&\\	
$0.088$  & $03.358$  & Sb	  &	&\\	
$0.0932$ & $22.1374$ & Sbc~I	  & 	&\\
$0.0939$ & $22.0434$ & S~IV	  & 	& \\
$0.1058$ & $14.1321$ & E0?	  &* 	&\\
$0.1112$ & $10.1281$ & S/Ir       & 	&\\		
$0.1178$ & $03.0443$ & Sb:	  & 	&\\	
$0.1409$ & $22.0676$ & Sa pec	  & 	& asymmetrical\\
$0.141 $ & $4.1281$  & Sa pec?	  &* 	&\\	
$0.1475$ & $10.0805$ & S t?	  & 	& \\	
$0.1510$ & $03.0500$ & S/Ir	  &*	& \\
$0.1550$ & $14.9025$ & Sbc pec	  &*	& \\
$0.1554$ & $03.1051$ & Sb~IV	  & 	&\\
$0.1877$ & $03.0332$ & Sbc:(+? + ?)& 	&\\	
$0.1952$ & $03.0982$ & Sa pec	   & 	&asymmetrical\\
$0.1955$ & $10.1653$ & Sbc: pec    & 	&\\
$0.1967$ & $10.1651$ & E5          & 	& \\		
$0.1970$ & $03.1014$ & Sc I        & 	& \\		
$0.1971$ & $10.1178$ & Sb          & 	&asymmetrical\\
$0.2000$ & $10.1161$ & E6/Sa       & 	& \\		
$0.2092$ & $14.1103$ & Compact	   &    &\\
$0.2187$ & $03.0365$ & Merger 	   & 	& \\		
$0.2220$ & $03.1387$ & E0/Star	   & 	& \\	
$0.223$  & $03.0315$ & Sb IV pec   &    &asymmetrical\\
$0.2230$ & $03.0494$ & 	?          &    &\\		
$0.2270$ & $14.0547$ & Merger:	   &*   &\\
$0.2342$ & $14.1209$ & S(B)b	   &*   &\\
$0.2345$ & $10.1643$ & S0/Sa	   &    & \\	
$0.2352$ & $14.1200$ & S	   &*   &\\
$0.2356$ & $14.1419$ & ...         &*   &image too small to classify\\
$0.2384$ & $14.0310$ & Sc pec	   &    & \\	
$0.2490$ & $22.0944$ & S(B)b I:	   &    & \\	
$0.2567$ & $14.1273$ & S pec	   &*   & \\	
$0.2596$ & $14.0377$ & Merger/Sc pec &  & \\	
$0.2640$ & $03.1050$ & SBc IV	   &    & \\	
$0.266 $ & $14.0695$ & Sa	   &    & \\	
$0.286 $ & $14.0983$ & Pec (tides) &*   & \\
$0.2902$ & $14.1242$ & E1 + ?      &    & \\		
$0.291$  & $14.1257$ & Sb	   &*   &edge-on\\
$0.2933$ & $22.0819$ & Pec	   &    & \\	
$0.2936$ & $22.0585$ & Sa	   &    & \\	
$0.2945$ & $22.0758$ & E1	   &    & \\	
$0.3090$ & $10.0802$ & Merger?	   &    & might also be part of a tidal 
tail\\
$0.3187$ & $22.0671$ & Sb	   &    & \\	
$0.3248$ & $14.0916$ & Sbc	   &*   & \\
$0.3251$ & $22.0622$ & Sb?	   &    & \\	
$0.3400$ & $10.0747$ & Sb (ring)   &    & \\		
$0.3410$ & $10.0831$ & Sa	   &    & \\	
$0.3481$ & $14.1446$ & E1	   &*   &halo of image contains a few blobs\\
$0.3595$ & $14.1071$ & Sb	   &*   &\\
$0.3600$ & $14.1503$ & E0	   &*   &\\
$0.3605$ & $03.0337$ & E1	   &    & \\	
$0.3616$ & $14.1239$ & Merger?	   &*   &\\
$0.3700$ & $03.0983$ & Sb:	   &    & \\	
$0.372$  & $14.0501$ & Sc	   &*   &\\
$0.3750$ & $14.1329$ & Sa pec	   &    & dusty\\
$0.3760$ & $03.0387$ & Sa	   &    & \\	
$0.3847$ & $10.0812$ & Sa	   &    & \\	
$0.420$  & $14.9987$ & Sb	   &*   & edge-on\\
$0.4210$ & $14.0422$ & E1	   &    & \\	
$0.4220$ & $03.1031$ & Sa	   &    & \\	
$0.4243$ & $22.0501$ & E0	   &    & \\	
$0.427$  & $14.1524$ & S(B)c	   &pec &*\\
$0.4300$ & $14.0998$ & Merger	   &*   &\\	
$0.4312$ & $22.0583$ & Sb pec	   &    & dusty, edge-on\\
$0.433$  & $14.1008$ & Sb	   &    &edge-on\\
$0.4345$ & $14.1179$ & E4/S0	   &*   &\\	
$0.4620$ & $14.1464$ & E3	   &*   &\\
$0.4640$ & $10.1250$ & SBab	   &    & \\	
$0.4640$ & $10.1259$ & E3	   &    & \\	
$0.4650$ & $10.1180$ & Sb          &    & \\		
$0.4667$ & $10.0813$ & Sb pec?	   &    & \\	
$0.4670$ & $10.1255$ & E0	   &    & \\	
$0.4681$ & $10.1349$ & Sa:	   &    & \\	
$0.4705$ & $22.0497$ & E3+E2+E2	   &    & perhaps early stage of triple 
merger\\
$0.4710$ & $10.1182$ & Ir	   &    & \\
$0.4733$ & $10.1231$ & Sa	   &    & \\	
$0.4738$ & $22.0919$ & Merger	   &    & \\	
$0.4750$ & $22.0609$ & Sb	   &    &edge-on\\
$0.4768$ & $22.0988$ & Ir	   &    & edge-on\\
$0.4788$ & $14.1012$ & S pec	   &*   & asymmetrical\\
$0.4800$ & $03.1060$ & S(B)ab	   &    & \\	
$0.4805$ & $03.0599$ & Sc IV	   &    & \\
$0.4825$ & $03.1373$ & Sb	   &    & \\	
$0.4870$ & $03.1413$ & Sab         &    & \\		
$0.4880$ & $03.1416$ & E0	   &    & \\
$0.4971$ & $10.1637$ & Sa	   &    & \\
$0.5070$ & $10.1155$ & Sb	   &    & asymmetrical\\
$0.5194$ & $10.1222$ & Sb pec      &    & \\		
$0.526$  & $10.0829$ & Merger	   &    & \\	
$0.5300$ & $03.0445$ & Sc          &    & \\		
$0.5301$ & $14.1395$ & Sc:	   &*    &\\
$0.5304$ & $03.0466$ & Ir?         &    &edge-on\\
$0.5460$ & $14.0207$ & E4          &    & \\		
$0.5489$ & $14.1037$ & ProtoSc?	   &*   &\\
$0.5500$ & $22.1037$ & Sa	   &    & \\	
$0.5520$ & $10.1153$ & SBbc:	   &    & \\	
$0.5620$ & $03.1347$ & S(B)(b?)    &    & \\		
$0.5773$ & $10.0793$ & Sbc: pec	   &    &asymmetrical, edge-on\\
$0.5775$ & $10.1262$ & E2	   &    & \\	
$0.5800$ & $10.0794$ & E1	   &    & \\	
$0.5820$ & $14.0725$ & Merger?	   &*   & \\
$0.5850$ & $10.1243$ & E3/Sa	   &    & \\	
$0.5940$ & $22.1279$ & Sa:	   &    & \\
$0.6016$ & $14.0393$ & Sc II	   &*   &\\
$0.6050$ & $03.1392$ & Sbc	   &*   & \\
$0.6056$ & $03.0485$ & Ir:	   &*   & edge-on\\
$0.6061$ & $03.0595$ & S pec	   &*   &asymmetrical, morphology quite 
different\\
         &           &             &    & in 450W and 814W\\
$0.6064$ & $03.0327$ & Sb:	   &*   &\\
$0.6069$ & $03.0488$ & Ir:	   &*   & \\
$0.6070$ & $03.0717$ & Sbc	   &*   &\\
$0.6079$ & $03.0480$ & Ir:	   &*   &\\
$0.6143$ & $14.0593$ & Sbt	   &*   &\\
$0.6180$ & $03.1032$ & E2	   & *  &\\	
$0.619$	 & $03.9003$ & Sc	   & *  &\\
$0.6200$ & $03.1319$ & E0 pec	   & *  & embedded in faint asymmetric 
nebulosity\\
$0.6232$ & $22.0453$ & Sab	   & *	&\\
$0.634$  & $14.0700$ & Sab	   & *	& \\
$0.6350$ & $03.1035$ & SBa	   & *	& \\
$0.6360$ & $03.1381$ & E3	   & *	& \\
$0.637$  & $03.1650$ & Sc/Ir	   &*	& \\
$0.6370$ & $14.0651$ & Sa	   &*	&\\
$0.6372$ & $03.1375$ & Sbc	   &*   &\\
$0.6404$ & $14.1136$ & S pec	   &*   &\\
$0.6410$ & $14.1043$ & S(B)b	   & *	&\\
$0.6430$ & $10.0826$ & Sc I	   &*   &\\	
$0.6449$ & $14.1258$ & E3	   &*	& \\
$0.6487$ & $10.1183$ & SB t	   &*	& \\
$0.6508$ & $03.0523$ & Merger	   &*   & \\	
$0.6545$ & $14.0485$ & Sc t	   &*   &\\
$0.6600$ & $03.579$  & Sc:	   &*	&\\
$0.6600$ & $14.1139$ & Merger	   &*	& \\
$0.6622$ & $14.0848$ & ?	   &*	&\\
$0.6690$ & $10.0769$ & Merger?	   &*	&\\
$0.6700$ & $10.1270$ & Sab/S0	   &*	&\\
$0.6710$ & $10.0763$ & Sbc II	  &*	&\\
$0.6710$ & $22.1078$ & Sa	&*	&\\
$0.671$  & $14.1164$ & Pec	&*	&\\
$0.6730$ & $14.1143$ & Sab	&*	&\\
$0.6740$ & $14.0400$ & Sc	&*	&\\
$0.6743$ & $14.0972$ & Pec	&*      &\\	
$0.6750$ & $14.0746$ & Sab	&*	&\\
$0.6758$ & $22.0945$ & Sab:	&*	&\\
$0.6862$ & $10.1203$ & Pec	&*	&\\
$0.690$	 & $03.1540$ & Sb t	&*	&\\
$0.6950$ & $03.0350$ & E3	&*	&\\
$0.6968$ & $03.0560$ & E2	&*	&\\
$0.703$  & $14.1264$ & ...	&*	&image too small to classify\\
$0.7040$ & $03.0999$ & Sc t	&*	&\\
$0.7054$ & $03.1016$ & S pec	&*	&off-center nucleus\\
$0.7062$ & $10.1207$ & Sa:	&*	&\\
$0.7080$ & $03.1395$ & Sb	&*	&\\
$0.7140$ & $03.0528$ & Sb/Sc:	&*	&\\
$0.715$  & $03.1531$ & S/Ir	&*	&\\
$0.7217$ & $14.1042$ & S pec	&*	&\\
$0.7242$ & $10.1423$ & E2	&*	&\\
$0.7310$ & $10.0817$ & Sc:	&*	&\\
$0.738$  & $10.811$  & S:	&*	&\\
$0.7426$ & $14.1126$ & Pec	&*	&\\
$0.7437$ & $14.1146$ & Sa + S Merger	&* &\\	
$0.7498$ & $10.1236$ & Sb	&*	&\\
$0.745$  & $14.1415$ & 	E + debris?	&*	&\\
$0.7526$ & $14.1189$ & Sb pec	&*	&\\
$0.7534$ & $14.0962$ &	S(B)a	&*	&\\
$0.7544$ & $14.1190$ &	Sc pec	&*	&\\
$0.7770$ & $10.1257$ & 	Sb	&*	&\\
$0.7850$ & $03.1384$ &  E0	&*	&\\
$0.7868$ & $10.0771$ &  Ir	&*	&\\
$0.8065$ & $14.1311$ &  cD:	&*	&\\
$0.8073$ & $14.0985$ & 	Merger	&*	&\\
$0.809$  & $14.0665$ & 	? + S: t	&*	&\\
$0.8100$ & $14.1277$ & 	E2 + debris	&*	&\\
$0.814$	 & $14.1251$ & 	Sb	&*	&\\
$0.8150$ & $03.0316$ &  Sb	&*	&\\
$0.8160$ & $10.1017$ &  Sbc:	&*	&\\
$0.8164$ & $22.1453$ &  Merger	&*	&\\
$0.8165$ & $10.1213$ & 	Sc:	&*	&\\
$0.818$  & $14.0749$ &	Pec	&*	&\\
$0.8182$ & $22.1406$ &	E:2	&*	&\\
$0.8191$ & $22.1313$  &	S t/Merger?	&*	&\\
$0.8194$ & $22.0764$  &	Sc? Pec		&*	&\\
$0.8270$ & $03.1499$  & Sb	&*	&\\
$0.8300$ & $22.0618$  &	E0	&*	&\\
$0.8307$ & $14.1356$  & Sb	&*	&\\
$0.831$  & $14.1129$  &	Merger	&*	&\\
$0.836$  & $14.0411$  &	Sb	&*	&\\
$0.8410$ & $10.1209$  &	E0	&*	&\\
$0.8520$ & $03.1393$  & Ir	&*	&edge-on\\
$0.8604$ & $14.1427$  &	SBb?	&*	&\\
$0.8750$ & $14.0899$  & Sa	&*	&\\
$0.8800$ & $03.0035$  & Sb pec	&*	&\\
$0.8891$ & $22.0599$  & Merger	&*	&\\
$0.8905$ & $22.0576$  & Merger	&*	&\\
$0.8990$ & $14.1496$  & E0	&*	&\\
$0.9011$ & $14.1079$  & Sc pec	&*	&\\
$0.9092$ & $10.1220$  &	Sb: t	&*	&\\
$0.918$  & $14.0939$  &	Merger	&*	&compact galaxy plus tidal debris\\
$0.9252$ & $22.0779$  & Sb:	&*	&\\
$0.9380$ & $03.1077$  &	Sab/cD	&*	&bright lensed arc\\
$0.9442$ & $03.1056$  & S pec / S t	&*	&\\
$0.9490$ & $10.1189$  &	Sab	&*	&\\
$0.9533$ & $22.1486$  & ...	&*	&image too small to classify\\
$0.969$  & $14.0608$  &	?	&*	&\\
$0.9728$ & $10.1615$  & S	&*	&\\
$0.9769$ & $22.0953$  & S pec /Merger	&*	&\\
$0.9832$ & $10.0761$  & Merger	&*	&\\
$0.985$  & $14.0807$  & ?	&*	&\\
$0.986$  & $14.0743$  & Sc pec	&*	&asymmetrical\\
$0.9876$ & $14.1028$  & Sa	&*	&\\
$0.9889$ & $14.0846$  & Merger?	&*	&\\
$0.9920$ & $14.0854$  &	Sb	&*	&\\
$0.9950$ & $14.9027$  &	Sc/Ir:	&*	&\\
$1.0151$ & $14.1166$  &	Sab:	&*	&\\
$1.0380$ & $03.1027$  & S pec	&*	&off-center nucleus\\
$1.0385$ & $14.0600$  &	Protogalaxy?	&*	&could also be ``tidal wreck''\\
$1.1810$ & $14.0147$  &	S pec /Ir	&&\\	
$1.1592$ & $10.1168$  &	Sc I:		&&\\
$1.1810$ & $14.0147$  &	S pec /Ir	&&\\	
$1.6034$ & $14.0198$  &	E1/Star         &&\\
\enddata
\end{deluxetable}

\clearpage

\begin{deluxetable}{lrrr}
\tablewidth{470pt}
\tablecaption{\sc{FREQUENCY DISTRIBUTION OF MORPHOLOGICAL TYPES IN CFRS SAMPLE 
\tablenotemark{a}}}

\tablehead{\colhead{Galaxytype} & \colhead{z = 0.25 -- 0.60} & \colhead{0.60 -- 
0.80} & \colhead{0.80 -- 01.20}}
\startdata

E + S0 + E/S0            & $22\%~~(11\%)$     &	 $15\%~~(21\%)$      & 
$10\%~~(16\%)$\\
E/Sa + S0/Sa             & $1~~~~~~(~0)$       &         $~0~~~(~0)$         & 
$~0~~~~(~2)$\\
Sa + Sab                 & $21~~~~~(15)$       &         $18~~~(11)$         & 
$~9~~~~(13)$\\
Sb + Sbc                 & $26~~~~~(26)$       &         $20~~~(23)$         & 
$22~~~~(10)$\\
Sc + Sc/Ir + Scd         & $9~~~~~~(10)$       &         $16~~~(~5)$         & 
$12~~~~(~5)$\\
Ir                       & $4~~~~~~(~5)$       &	 $~7~~~(~7)$         & 
$~3~~~~(12)$\\
S	                 & $3~~~~~~(11)$       &	 $10~~~(10)$         & 
$~9~~~~(11)$\\
Pec. + ? + Protogalaxy   & $4~~~~~~(16)$       &         $~9~~~(19)$	     & 
$15~~~~(12)$\\
Merger                   & $9~~~~~~(~7)$       &         $~5~~~(~4)$         & 
$18~~~~(15)$\\
Total in sample  	 & $72~~~~~(50)$       &         $64~~~(70)$	     & 
$49~~~(120)$\\

\enddata
\tablenotetext{a}{For CFRS, with previous results for HD + FF in parentheses}
\end{deluxetable}

\clearpage

\begin{deluxetable}{lrrr}
\tablewidth{470pt}\tablecaption{\sc{MORPHOLOGICAL TYPES IN COMBINED CFRS + HST 
SAMPLES \tablenotemark{a}}}
\tablehead{\colhead{Galaxy type} & \colhead{z = 0.25 -- 0.60} & \colhead{0.60 -- 
0.80} & \colhead{0.80 -- 1.20}}
\startdata

E + S0 + E/S0	       & $21\%~~(17\%)$     &  $24.5\%~~(18\%)$    & 
$~25\%~~(15\%)$\\
E/Sa + S0/Sa	       & $~~1~~~~(~1\%)$    &  $~~0~~~~~~(~0\%)$   & 
$~~3~~~~(~2\%)$\\
Sa + Sab	       & $~23~~~~(19\%)$    &  $~19.5~~~~(15\%)$   & 
$~20.5~~(12\%)$\\
Sb + Sbc	       & $~32~~~~(26\%)$    &  $~29~~~~~~(22\%)$   & 
$~23~~~~(14\%)$\\
Sc + Sc/Ir + Scd       & $~11.5~~(~9\%)$    &  $~13~~~~~~(10\%)$   & 
$~12~~~~(~7\%)$\\
Ir	               & $~~5~~~~(~4\%)$    &  $~~9.5~~~~(~7\%)$   & 
$~16.5~~(10\%)$\\
S	               & $~~8~~~~(~7\%)$    &  $~13.5~~~~(10\%)$   & 
$~18.5~~(11\%)$\\ 
Pec. + ? + Protogalaxy & $~~1~~~~(~9\%)$    &  $~19~~~~~~(14\%)$   & 
$~22.5~~(13\%)$\\
Merger	               & $~~9.5~~(~8\%)$    &  $~~6~~~~~~(~4\%)$   & 
$~28~~~~(17\%)$\\
Total in sample	      &~$122~~~~~~~~~$   &~~~~$134~~~~~~~~~~~$  
&~~~~$169~~~~~~~~~$\\

\enddata
\tablenotetext{a} {Due to rounding erros not all percentages add up to 100\% }
\end{deluxetable}

\clearpage

\begin{figure}
\caption{Frequency distribution of classification differences in the sense 
Binchmann et al. minus the present classifications reduced to the Binchmann 
numerical system.  The figure shows that 54\% of the classifications differ by 
only 0.5 classification bins or less.}
\end{figure}

\begin{figure}
\caption{The galaxy CFRS 14.1129 is classified as an irregular by Binchmann et 
al., but appears to be a late-type spiral that is being tidally distorted by an 
early-type galaxy.}
\end{figure}

\begin{figure}
\caption{Frequency distribution of redshifts in the present sample.  The numbers 
per redshift bin are seen to increase slowly up to $z=0.9$ or $z=1.0$.  The drop 
beyond $z=1.0$ is due to incompleteness.}
\end{figure}

\begin{figure}
\caption{Comparison between the redshift distribution of the spiral and 
elliptical galaxies in the CFRS sample.  A Kolmogorov-Smirnov test shows that 
the differences between these two distributions are not statistically 
significant.}
\end{figure}


\clearpage

\begin{thebibliography}{}

\bibitem [Abraham \etal\ (1999)] {abrahametal1999} Abraham, R. G., Merrifield, 
M. R., Ellis, R. S., Tanvir, N. R., \& Binchmann, J. 1999, \mnras, 308, 569

\bibitem [Binchmann \etal\ (1998)] {binchmannetal1998} Binchmann, J., 1998, 
\apj, 499, 112

\bibitem [Carlberg \etal\ (2000)] {carlbergetal2000} Carlberg, R. G., 2000, 
\apj, 532, L1

\bibitem [Carollo \& Lilly (2001)] {carollolilly2000} Carollo, C.M., \& Lilly, 
S.J. 2001, \apj, 548, L153

\bibitem [Crampton \etal\ (1995)] {crampton1995} Crampton, D., Le F\`{e}vre, O., 
Lilly, S. J., \& Hammer, F. 1995, \apj, 455, 96

\bibitem [Dressler \etal\ (1980)] {dressler1980} Dressler, A. 1980, \apj, 236, 
351-365

\bibitem [Eskridge \etal\ (2000)] {eskridgeetal2000} Eskridge, P. B.  2000, \aj, 
119, 536

\bibitem [Ferguson (2000)] {ferguson2000} Ferguson, H. C., Dickinson, M., \& 
Williams, 2000, \araa, 38 in Press (2000) (=astro-ph/004319)

\bibitem [Hammer \etal\ (1995)] {hammer1995} Hammer, F., Crampton, D., Le 
F\`{e}vre, O., \& Lilly, S. J. 1995, \apj, 455, 88

\bibitem [Hubble (1936)] {hubble1936} Hubble, E. 1936, The Realm of the Nebulae 
(Yale Univ. Press, New Haven), pp. 36-57

\bibitem [Le F\`{e}vre \etal\ (1995)] {LeFevreetal1995}  Le F\`{e}vre, O., 
Crampton, D., Lilly, S.J., Hammer, F., \& Tresse, L. 1995, \apj, 455, 60

\bibitem [Lilly \etal\ (1995a)] {lilly1995a} Lilly, S. J., Le F\`{e}vre, O., 
Crampton, C., Hammer, F. \& Tresse, L., 1995a, \apj, 455, 50

\bibitem [Lilly \etal\ (1995b)] {lilly1995b} Lilly, S.J., Hammer, F., Le 
F\`{e}vre, O., \& Crampton, D. 1995b, \apj, 455, 75

\bibitem [Sandage (1981)] {sandagetammann1981} Sandage, A. \& Tammann, G. A. 
1981, A Revised Shapley-Ames Catalog of Bright Galaxies (Carnegie Institution:  
Washington), p. 91

\bibitem [van den Bergh (1960a)] {vdb1960a} van den Bergh, S. 1960a, \apj, 131, 
215

\bibitem [van den Bergh (1960b)] {vdb1960b} van den Bergh, S. 1960b, \apj, 131, 
558

\bibitem [van den Bergh (1960c)] {vdb1960c} van den Bergh, S. 1960c, Pub. David 
Dunlap Obs. Vol. 2, No. 6

\bibitem [van den Bergh (1998)] {vdb1988} van den Bergh, S. Galaxy Morphology 
and Classification, 1998 (Cambridge Univ. Press: Cambridge)

\bibitem [van den Bergh \etal\ (1996)] {vdb1996} van den Bergh, S., Abraham, 
R.G., Ellis, R., Tanvir, N.R., Santiago, B. X., \& Glazebrook, K. 1996, \aj, 
112, 359

\bibitem [van den Bergh, Cohen, \& Crabbe (2001)] {vdbcohencrabbe2001} van den 
Bergh, S., Cohen, J.C., \& Crabbe, F. 2001, \aj (in press)

\bibitem [van den Bergh \etal\ (2000)] {vdb2000} van den Bergh, S., Cohen, J. 
G., Hogg, D.W., \& Blandford, R. 2000, \aj, 120, 2190

\bibitem [Visvanathan \& van den Bergh (1992)] {visvanathanvdb1992} Visvanathan, 
N., \& van den Bergh, S. 1992, \aj, 103, 1057

\bibitem [Williams \etal\ (1996)] {williamsetal1996} Williams, R. E.  1996, \aj, 
112, 1335 

\end{thebibliography}
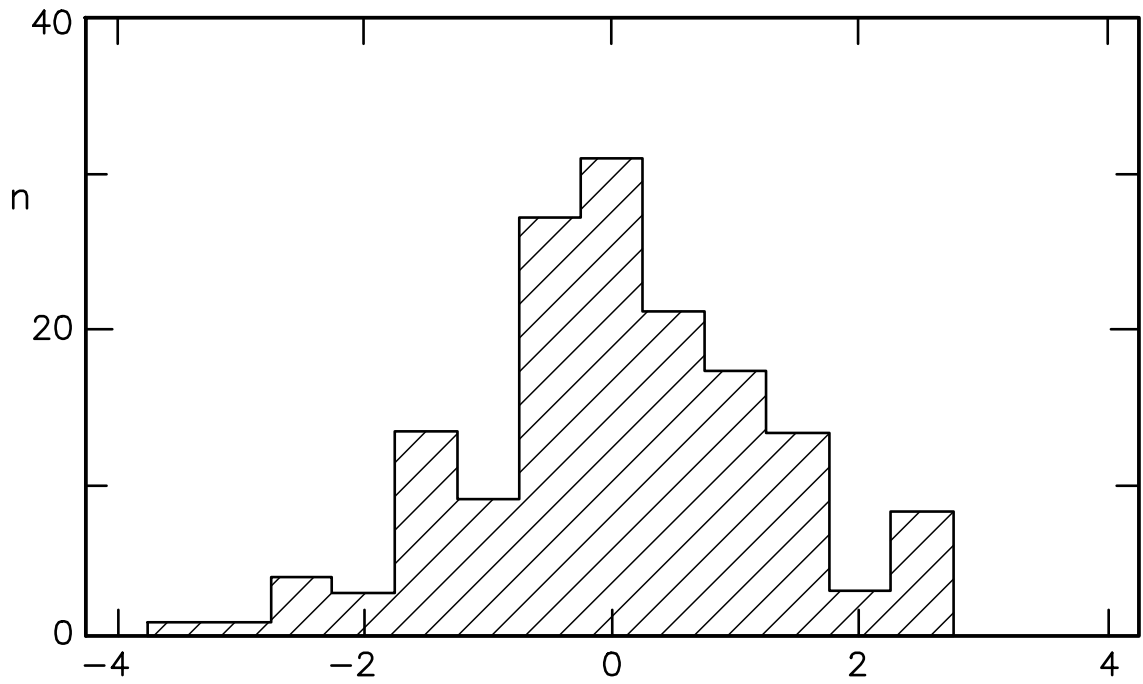
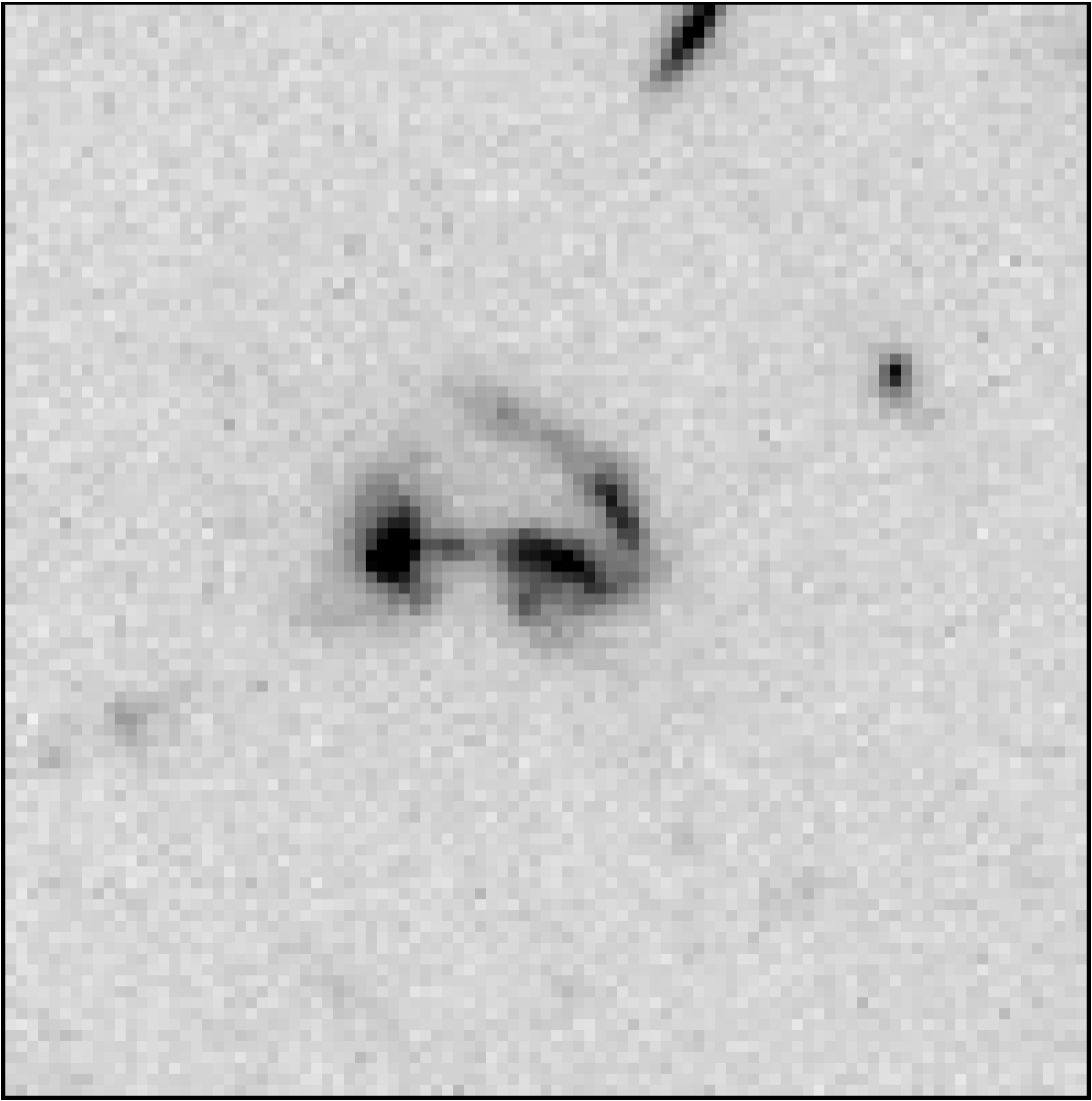
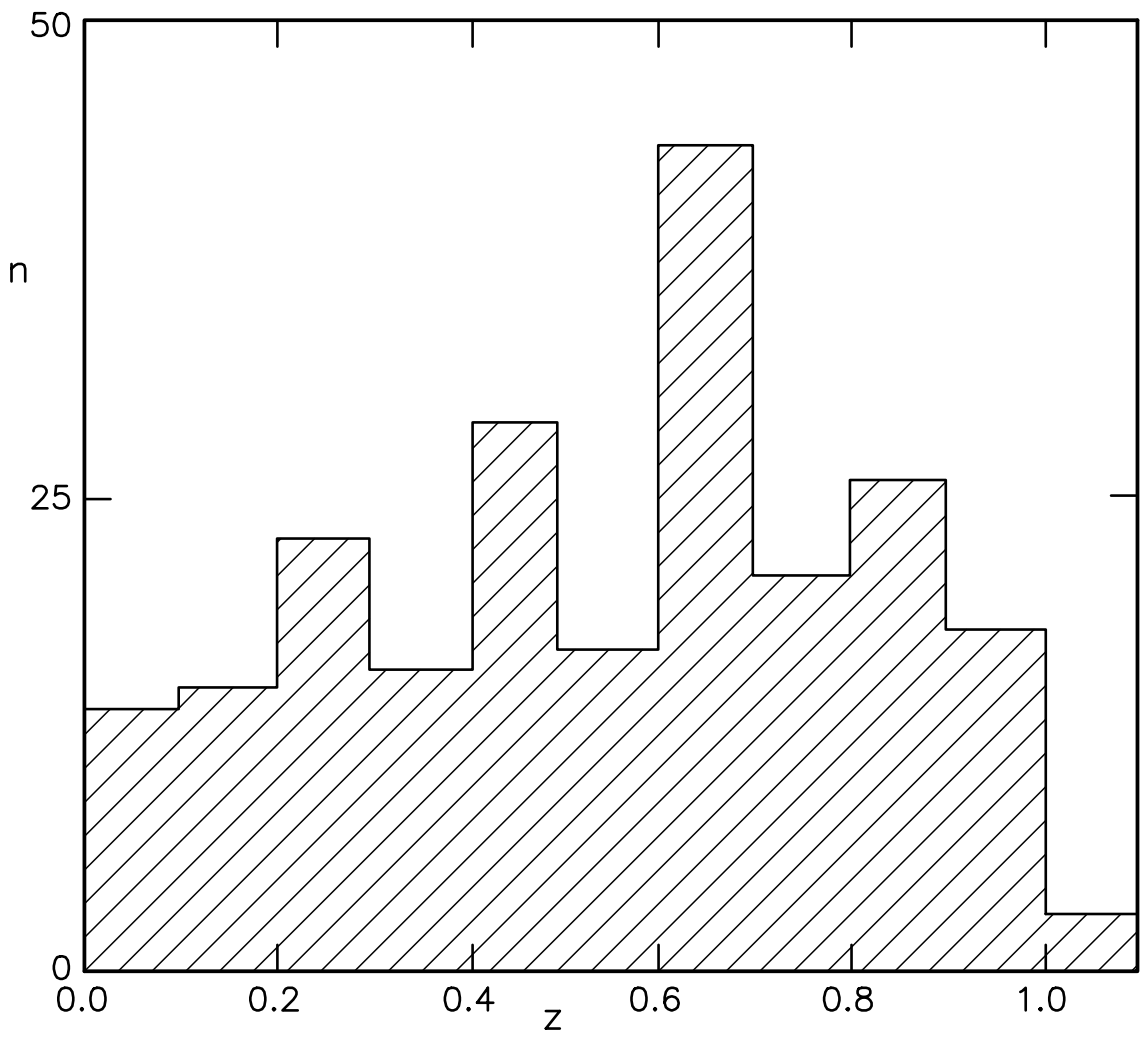
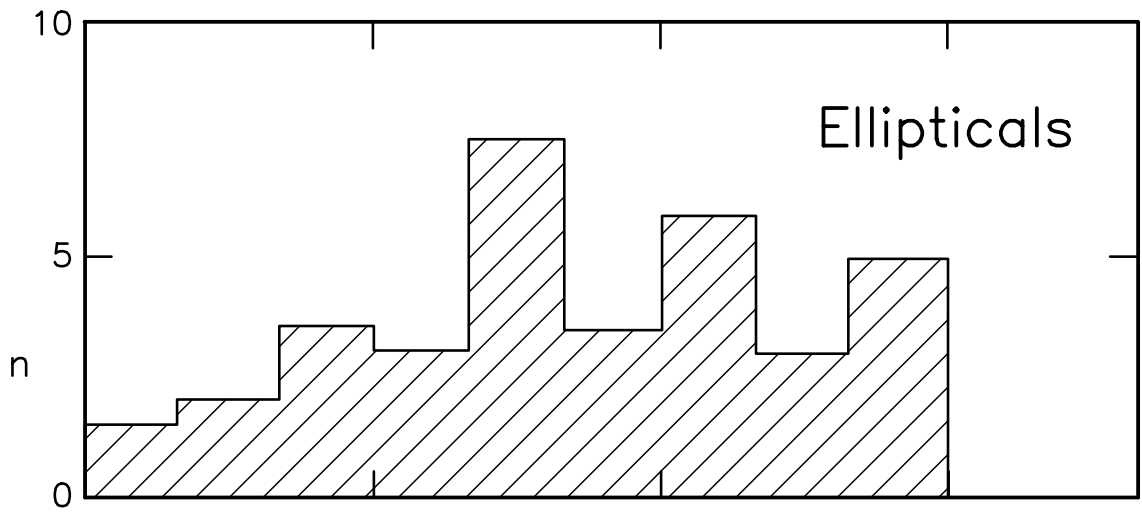
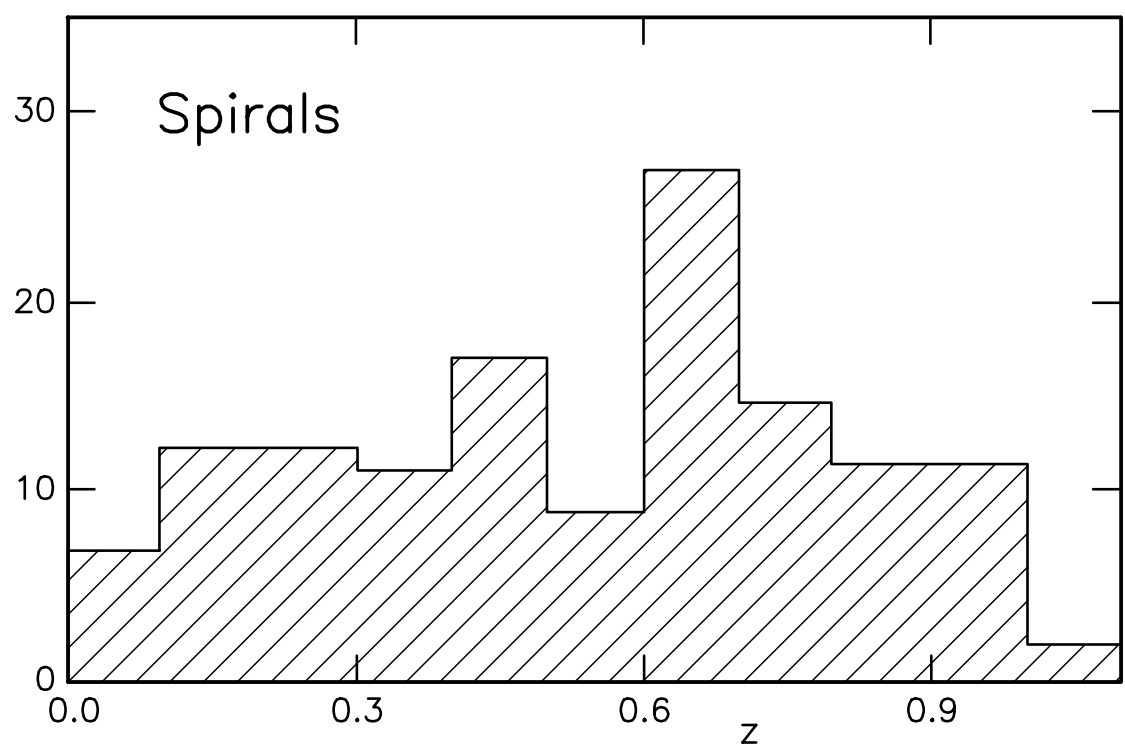
\end{document}